\documentclass[12pt]{article}
\begin{document}
\mbox{}
\vspace{0.5cm}
\begin{center}
\LARGE
\textbf{The epistemic view of quantum states
and the ether}\\[1cm]
\large
\textbf{Louis Marchildon}\\[0.5cm]
\normalsize
D\'{e}partement de physique,
Universit\'{e} du Qu\'{e}bec,\\
Trois-Rivi\`{e}res, Qc.\ Canada G9A 5H7\\
email: marchild$\hspace{0.3em}a\hspace{-0.8em}
\bigcirc$uqtr.ca\\
\end{center}
\medskip
\begin{abstract}
The idea that the wave function represents
information, or knowledge, rather than the state
of a microscopic object has been held to solve
foundational problems of quantum mechanics.
Realist interpretation schemes, like Bohmian
trajectories, have been compared to the ether
in pre-relativistic theories.  I argue that the
comparison is inadequate, and that the epistemic
view of quantum states begs the question of
interpretation.
\end{abstract}
\medskip
\textbf{PACS Nos.:} 03.65.Ta, 03.50.De, 03.30.+p
\section{Introduction}
For the past twenty years, quantum information
theory has been one of the most rapidly developing
areas of quantum mechanics~\cite{nielsen}.
Teleportation of unknown quantum states over large
distances has been reported, and practical
implementations of quantum cryptography are already
available.  Although large-scale quantum computers
may not be coming soon, work in that direction is
likely to throw much light on the fundamental process
of decoherence and perhaps on the limits of quantum
mechanics itself.

Investigations in the foundations of quantum
mechanics have significantly contributed to
the development of quantum information theory.
A number of people believe that in turn,
quantum information theory has much to contribute
to the understanding and interpretation of quantum
mechanics.  This has to do with what is often
called the \emph{epistemic view} of quantum states,
which goes back at least to Heisenberg but has
significantly evolved in the past few
years~\cite{rovelli,peres,fuchs}.
The basic assertion of the epistemic view is that
the wave function (or state vector, or density matrix)
represents knowledge, or information.  On what
the wave function is knowledge of, proponents of
the epistemic view do not necessarily agree.
The variant most relevant to the present discussion is
that rather than referring to objective
properties of microscopic objects (such as
electrons, photons, etc.), the wave function
encapsulates probabilities of results
of eventual macroscopic measurements.

The proponents of the epistemic view believe that
it considerably attenuates, or even completely
solves, the notorious problems of quantum measurement
and long-distance correlations.  Briefly, if the wave
function is interpreted as referring to the objective
state of a physical system, its collapse in a quantum
measurement involves a physical process that calls for
explanation.  If, however, the wave function simply
represents knowledge of the probabilities of results,
its abrupt change points to a change of knowledge,
rather than to a physical change in some microscopic
system.  In a similar way, the epistemic view helps removing
the clash between collapse and Lorentz invariance.
In the EPR setup, for instance, Alice's measurement
of her photon's spin does not instantaneously
produce the collapse of
the spin wave function of Bob's photon.  Rather it changes
Alice's knowledge of the probabilities of results of spin
measurements on Bob's photon.  Bob can of course perform the
measurement, but otherwise Alice's knowledge can only
be transferred to him by conventional means. 

In the epistemic view, the Hilbert space formalism
of quantum mechanics is taken as complete, and its
objects in no need of a realistic interpretation.
Thus any additional constructs, like value assignments
in modal interpretations~\cite{vermaas,fraassen},
multiple worlds~\cite{everett}, or Bohmian
trajectories~\cite{bohm1,bohm2,holland}
are viewed as superfluous at best.  Such constructs
predict no empirical consequences other than what is already
derivable from the Hilbert space formalism.  This has
lead to comparing them with the ether in classical
electrodynamics~\cite{bub1,bub2,bohr}.  Just as the
ether was discarded after special relativity had shown
that it led to no specific empirical consequences, so
should additional constructs to the Hilbert space be
done away with in quantum mechanics.

The purpose of this paper is to investigate whether
the roles fulfilled by the ether in electrodynamics
and by realistic interpretation schemes in quantum mechanics
are comparable, and whether the epistemic view provides
an adequate understanding of quantum
states.\footnote{Further thoughts on these questions
can be found in Refs.~\cite{marchildon1}
and~\cite{marchildon2}, on
which the present discussion is based.}
\section{Ether and field}
The concept of ether has a long history, and it has
been the subject of detailed analyses~\cite{whittaker,
darrigol}.  In one of its many uses, it was viewed as
a substratum wherein electric and magnetic phenomena
take place.  In the nineteenth century, a number of
complicated mechanical models were proposed to
account for its properties, by such distinguished
physicists as J.~C.~Maxwell and W.~Thomson, among
others.  Although a review of such models would lead
us too far astray, it is appropriate to look at the
way the ether was seen by H.~A.~Lorentz, at the turn
of the twentieth century.

By that time Lorentz had settled on the concept of a
stationary ether introduced by A.~J.~Fresnel in the
first decades of the nineteenth century.  Materiel
bodies moving through the ether would leave it
undisturbed, except for carrying their own excess
ether responsible for such characteristics as their index
of refraction.  The stationary ether explained very well
phenomena like stellar aberration.  It was viewed as
defining an absolute reference frame, the one in
which Maxwell's equations would hold.

Lorentz's point of view on electromagnetic theory,
``a surprising and audacious step'' in the words of
Einstein~\cite[p.~35]{einstein}, was much simpler
than his predecessors'.  For him there is ponderable
matter, there are electric charges (``electrons''), and
there is the ether, ``the receptacle of electromagnetic
energy and the vehicle for many and perhaps for all
the forces acting on ponderable
matter''~\cite[p.~30]{lorentz}.  Lorentz did not commit
himself on whether all matter is made of electric
charges, nor on whether all mass has an electromagnetic
origin.

As there was no reason to expect that the earth is at
rest with respect to the ether, the question arose as
to how to describe electromagnetic phenomena in a moving
frame.  What Lorentz and Poincar\'{e} progressively
realized was that one can define coordinates associated
with the moving frame and linear combinations of the
electric and magnetic field components such that
these combinations and coordinates satisfy Maxwell's
equations.  This was interpreted as meaning that
no electromagnetic (and, in particular, optical)
measurements could reveal motion through the ether.
Yet Lorentz and Poincar\'{e} still appealed to a
dynamical deformation of bodies in motion, and
retained the notions of absolute rest and absolute
time.  It was Einstein's fundamental contribution
to see that these notions could be dispensed with,
and that all inertial frames were equivalent in
all respects.

Einstein could therefore do away with the ether.
But the field concept became all the more important.
Hitherto dependent on the ether, the field now acquired
its full autonomy.  It was no longer viewed as in
need of a substratum.  The rejection of the ether,
therefore, did not leave a void in its stead.  The
ether was discarded as defining absolute time and
motion, but its function with respect to
electromagnetic phenomena was transferred to the
field itself.
\section{Bohmian trajectories}
Bohm's approach to quantum mechanics is an example
of a realistic interpretation.  It can be formulated
equally within the context of the Schr\"{o}dinger
or the Dirac equation.  We shall look more
specifically at the latter.

The Dirac equation for a particle of mass $m$
and charge $e$ in an external electromagnetic field
$A^{\mu}$ is given by
\begin{equation}
-i \gamma_{\mu} \frac{\partial \psi}{\partial x_{\mu}}
+ \frac{e}{\hbar} \gamma_{\mu} A^{\mu} \psi
+ \frac{mc}{\hbar} \psi = 0 . \label{dirac}
\end{equation}
Eq.~(\ref{dirac}) implies the existence of a
conserved current $j_{\mu} = \bar{\psi} \gamma_{\mu} \psi$,
where $\bar{\psi} = \psi^{\dagger} \gamma_0$.
Bohmian trajectories can be introduced by
specifying that the three-velocity of the particle at
the space-time point $(\mathbf{r}, t)$ is given by
\begin{equation}
\mathbf{v} = c \frac{\mathbf{j}}{j_0} .
\label{velocity}\end{equation}
It can be shown~\cite{bohm2,holland}
that the magnitude of the velocity never exceeds $c$,
and that if particles are distributed according
to the probability density $j_0 = \psi^{\dagger} \psi$
at a given time and follow the trajectories,
they will be distributed according
to $\psi^{\dagger} \psi$ at any other time.  Averages
computed on an ensemble of Bohmian particles exactly
coincide with averages computed by means of the Dirac
equation.

Thus Bohmian trajectories make no empirical predictions
not already obtainable from standard (Schr\"{o}dinger
or Dirac) quantum mechanics.  A similar remark can be
made about all realistic interpretations of quantum
mechanics that leave its basic formalism intact.
Just like the (unobservable) ether was discarded
with the advent of special relativity, shouldn't
we do away then with the trajectories or other additional
constructs to the Hilbert space structure?

Before examining this question, one more analogy between
quantum mechanics and the ether should be pointed out.
We have seen that in the usual approach to quantum
measurement, the collapse of the wave function is
essentially instantaneous.  Alice's measurement of her
photon immediately produces the collapse
of the wave function of Bob's.
It would then seem that wave function collapse
introduces a preferred reference frame.  Such a frame
also appears to be required in the Bohmian mechanics
of many particles~\cite{bohm2,maudlin}.  One can show that
this does not prevent the construction of a relativistically
covariant theory of \emph{observables}, but it is a strong
obstacle to the construction of such a theory
of \emph{beables}~\cite{bell}.
\section{Two explanatory roles}
Bohmian trajectories and the ether are elements
of two different theoretical structures.  They
present both analogies and differences.  The
analogy that is relevant here is that neither Bohmian
trajectories in quantum mechanics nor the ether in
special relativity lead to specific empirical
consequences.  Does this mean that the trajectories,
or other interpretative devices,
have in quantum mechanics the same status as the
ether in special relativity?  And if one can
dispense with such devices, is there
something which, like the field, plays the
role they would otherwise have?

To examine these questions, it is appropriate
to start with the following observation.
Although all measurements are made
by means of macroscopic apparatus, quantum
mechanics is used, as an explanatory theory,
in two different ways: it is meant to explain
(i) nonclassical correlations between macroscopic
objects and [ultimately through quantum field
theory] (ii) the small-scale structure of
macroscopic objects.  To show that these two functions
are distinct, we will consider a hypothetical
situation where only one of them is operating.

Consider a world where objects that are not too small
(say, larger than the wavelength of visible light)
behave, for all practical purposes, like similar
objects in the real world.  Classical mechanics can
be used to compute the trajectories of projectiles,
and classical hydrodynamics the flow of water in pipes.
Antennas and waveguides are described by 
Maxwell's equations.  Chemical equilibrium and
phase transitions obey the laws of
classical thermodynamics.  All objects in the
solar system have trajectories well described by
Newton's laws of gravitation and motion, perhaps
slightly corrected by the equations of
general relativity.

As we go down to scales much smaller than a fraction
of a micron, however, these laws
may no longer hold.  Except for one restriction
soon to be spelled out, I shall not be specific
about the changes that macroscopic laws may or
may not undergo in the microscopic realm.  Matter,
for instance, could either be continuous down to
the smallest scales, or made of a small
number of constituent particles like our
atoms.  The laws of particles and fields
could be the same at all scales, or else
they could undergo significant changes as smaller
and smaller distances are being probed.

In the hypothetical world, some macroscopic
objects at times behave in ways that cannot be
explained by the classical theories.  There may be,
for example, objects like our Geiger counters that
click when objects like our radioactive materials
are brought nearby.  Or there may be instruments
like our Stern-Gerlach devices which, when placed
in front of an oven and suitable collimators, modify
the pattern of blackening on a plate behind.  There
may even be large objects like some of our particle
accelerators, which in appropriate situations
produce various tracks in saturated vapour.  In
all these instances, the probabilities of
occurrence of events can be calculated on the
basis of the quantum-mechanical rules.

To explain the nonclassical correlations described
above between macroscopic objects,
one can think of at least two very different conceptual
schemes.  One can assume the existence of microscopic
objects, or ``particles,''
going from emitters to detectors, ovens to
plates or accelerators to vapour, whose properties
correspond to operators in the Hilbert spaces
used to compute the probabilities.  Or one can refrain
from postulating such microscopic objects, and assume
instead something like \emph{genuine
fortuitousness}~\cite{bohr,ulfbeck}, where clicks
or ``detection events'' are essentially uncaused.
\section{Only one world}
I now make an assumption about the hypothetical world, 
which characterizes the fundamental
way in which it differs from
the world we live in.  I suppose that the ``particles''
used in the first explanatory scheme above have nothing
to do with the microscopic structure of the
macroscopic objects.  That is, the ultimate constituents
of matter, if any, are completely different from
whatever is responsible for the nonclassical
correlations.  In the hypothetical
world, the function of the quantum-mechanical rules
is solely to explain these nonclassical correlations.
In that case the explanatory schemes of
particles and genuine fortuitousness are both
adequate to the job.

In the actual world, however, the situation is
different.  The observables, quantum numbers, and Hilbert
spaces relevant to the description of particles
responsible for macroscopic correlations are the same
as the ones used in describing the microscopic
structure of macroscopic objects.  Rutherford's
$\alpha$ particles produced by radioactive radon and
scattered by thin foils of gold have the quantum
numbers of helium nuclei!

How are proponents of the epistemic view
going to deal with that?  One way is to
adopt a strong instrumentalist stance and deny either
that microscopic objects exist or have states.
Indeed~\cite[p.~410]{bohr}

\begin{quote}
[i]t is a hallmark of the theory based on genuine
fortuitousness that it does not admit physical variables.
It is, therefore, of a novel kind that does not deal
with things (objects in space), or measurements, and
may be referred to as the theory of no things.
\end{quote}

The view that microscopic objects do not exist or have no
states is not logically inconsistent.  But it raises
the question, How can macroscopic objects exist and
have states, and yet be reducible to microscopic objects
that either do not exist or do not have states?  
How can the world be for the formalism of quantum
mechanics to be true?  That question can in fact be viewed
as the fundamental problem of the interpretation of
a theory~\cite{fraassen}.  Interpreting a theory
coincides with giving one (or several) ways in which
the formalism of the theory can be truly realized.
The upshot is that the strong instrumentalist stance
that microscopic objects don't exist or don't have
states does not constitute an interpretation, but
asks for one.

Without going so far as denying the existence of
microscopic objects or their states, proponents
of the epistemic view can claim that their
introduction is methodologically
inappropriate~\cite[p.~260]{bub1}.  
\begin{quote}
[I]f $T'$ and $T''$ are empirically equivalent
extensions of a theory $T$, and if $T$ entails that,
in principle, there \emph{could not be} evidence
favoring one of the rival extensions $T'$ or $T''$,
then it is not rational to believe either $T'$ or $T''$.
\end{quote}
Here $T$ can stand for the Hilbert space formalism of
quantum mechanics, $T'$ for its Bohmian extension,
$T''$ for Everett's worlds, etc.  If $T$ is singled
out among its empirical equivalents, it must be
on the basis of criteria other than empirical,
perhaps something like Ockham's razor.  This comes
as no surprise since even within the class of
internally consistent theories, acceptance almost
never depends on empirical criteria alone.

But here $T$ is just not
complete.  The Hilbert space of quantum mechanics
makes contact with experiments by means of
ill-defined concepts, like the one of a
macroscopic apparatus.  We are never told what
precise criteria of size, mass or constitution
make an aggregate of matter an apparatus.
$T'$ and $T''$ may be preferable to $T$ just
because they are more complete.

Neither Bohmian trajectories nor the ether lead
to specific consequences.  There have been
suggestions that just as the ether was replaced
by the field with the advent of special relativity,
additional constructs to the Hilbert space
formalism should be discarded and replaced by
the emergence of the concept of information.
\begin{quote}
[J]ust as Einstein's analysis (based on the
assumption that we live in a world in which
natural processes are subject to certain
constraints specified by the principles of
special relativity) shows that we do not need
the mechanical structures in Lorentz's theory
(the aether, and the behaviour of electrons in
the aether) to explain electromagnetic phenomena,
so the CBH analysis
(based on the assumption that we live in a world
in which there are certain constraints on the
acquisition, representation, and communication
of information) shows that we do not need the
mechanical structures in Bohm's theory (the guiding
field, the behaviour of particles in the guiding
field) to explain quantum phenomena~\cite[p.~262]{bub1}.
\end{quote}

The CBH analysis referred to is an important result
recently obtained by Clifton, Bub, and
Halvorson~\cite{clifton,halvorson}.
Working in the setting of $C^*$-algebras, these
investigators characterized the quantum theory by three
properties: (i) kinematic independence,
i.e.\ the commutativity of
the algebras of observables pertaining to distinct
physical systems; (ii) the noncommutativity of
an individual system's algebra of observables;
and (iii) nonlocality, i.e.\ the existence of entangled
states for spacelike-separated systems.  They
then showed that these properties are equivalent to
three information-theoretic constraints, namely,
the impossibility of superluminal information
transfer, of perfect broadcasting, and of
unconditionally secure bit commitment.

The concept of information is no doubt relevant
to the first explanatory function of quantum
mechanics, the one that pertains to nonclassical
correlations between macroscopic objects.  But
it is of no help in accounting for the microscopic
structure of macroscopic objects.  No proponent
of the epistemic view (as far as I know) would
go so far as claiming that objects are made of
information.  This is in sharp contrast with the
concept of field which, unlike information,
does not need material support and carries energy
and momentum of its own.  Even in classical
electrodynamics, there were proposals that all the
mass of charged particles is in fact field
energy~\cite{mccormmach}.  The analogy between
field and information is defective in an essential way.
\section{Conclusion}
Neither the ether nor Bohmian trajectories have
specific empirical consequences.  Special relativity,
while rejecting the ether as defining an absolute
reference frame, transferred its function of substantive
medium to the field itself.  Bohmian trajectories, or
other interpretative schemes of quantum mechanics, try
to make the basic variables of the theory, in terms of
which the structure of macroscopic objects is ultimately
explained, intelligible.  This makes them fundamentally
different from the ether, and points to the inadequacy
of the epistemic view of quantum states.
\end{document}